\documentclass[aps,prb, twocolumn,superscriptaddress]{revtex4-2}

\usepackage{graphicx}
\usepackage[dvips]{color}
\usepackage{color} 

\begin{document}


\title{Stabilization of nonmagnetic hexagonal close-packed phase in nanocrystalline nickel metal with a giant volume expansion}



\author{Sho Otsuru}
\affiliation{Department of Physics, Saga University, Saga 840-8502, Japan}
\author{Kenta Akashi}
\affiliation{Department of Applied Quantum Physics, Kyushu University, Fukuoka 819-0395, Japan}
\author{Miki Kakihara}
\author{Takuto Tsukahara}
\affiliation{Department of Physics, Saga University, Saga 840-8502, Japan}

\author{Hirofumi Ishii}
\author{Yen-Fa Liao}
\author{Ku-Ding Tsuei}
\affiliation{National Synchrotron Radiation Research Center, Hsinchu 30076, Taiwan}

\author{Yuji Inagaki}
\author{Tatsuya Kawae}
\affiliation{Department of Applied Quantum Physics, Kyushu University, Fukuoka 819-0395, Japan}

\author{Tetsuya Kida}
\affiliation{Department of Applied Chemistry and Biochemistry, Kumamoto University, Kumamoto 860-8555, Japan}

\author{Satoshi Suehiro}
\affiliation{Materials Research and Development Laboratory, Japan Fine Ceramic Center (JFCC), Nagoya 456-8587, Japan }

\author{Masashi Nantoh}
\affiliation{Advanced Device Laboratory, The Institute of Physical and Chemical Research (RIKEN), Wako 351-0198, Japan}

\author{Koji Ishibashi}
\affiliation{Advanced Device Laboratory, The Institute of Physical and Chemical Research (RIKEN), Wako 351-0198, Japan}
\affiliation{CEMS, RIKEN, Wako 351-0198, Japan}

\author{Yoichi Ishiwata}
\email{ishiwata@cc.saga-u.ac.jp}
\affiliation{Department of Physics, Saga University, Saga 840-8502, Japan}



\date{\today}

\begin{abstract}
Nanocrystalline Ni can adopt a hexagonal close-packed (hcp) structure,  
which does not appear in a bulk form under any conditions. 
Structural studies found its $>$ 20\% expansion of volume per atom compared with face-centered cubic (fcc) Ni.   
Hard x-ray photoemission spectroscopy of the hcp phase revealed a $>$ 0.14 valence electron transfer from 3$d$ to 4$s$ and 4$p$ orbitals 
and a 0.25 eV reduction in band energy per electron compared with the fcc phase. 
Its nonmagnetic nature was attributed to a low density of states at the Fermi level.  
Thermodynamic stability of the hcp phase can be explained by the band energy gain and the electrostatic potential energy loss 
as related to the large lattice expansion.
\end{abstract}

\pacs{}

\maketitle


The crystal structure of transition metals determines their band structure,  
which directly influences their resulting physical properties such as magnetism. 
The determination of a thermodynamically stable structure has a strong correlation to the $d$-occupation number ($n_d$) \cite{Skriver, Pettifor}. 
In the series of 4$d$ metals, Y and Zr form hexagonal close-packed (hcp) structures, 
Nb and Mo have body-centered cubic (bcc) structures, 
Tc and Ru adopt hcp structures, 
and finally Rh and Pd exhibit face-centered cubic (fcc) structures---this hcp-bcc-hcp-fcc crystal structure sequence is reproduced for 5$d$ metals. 
The 3$d$ metal series also has this hcp-bcc-hcp-fcc order, 
although the magnetic 3$d$ metals (including Mn, Fe and Co) exhibit different structures. 
This is because the magnetic energy contribution is much greater than the structural band energy difference \cite{Hasegawa}, 
so the structure with a high density of states (DOS) peak in the vicinity of the Fermi level ($E_F$), 
satisfying the Stoner criterion, 
is predominantly stabilized if exists.

The application of pressure to a ferromagnetic 3$d$ metal causes broadening of the $d$ bands and also $s$-$d$ charge transfer 
which increases $n_d$ because of the difference in spatial extension between the orbitals \cite{McMahan}. 
These effects reduce the DOS at $E_F$ and can cause a structural phase transition to a nonmagnetic phase 
with the same structure as their isoelectronic 4$d$ and 5$d$ counterparts. 
Accordingly, Fe converts from a ferromagnetic bcc to a nonmagnetic hcp phase（matching Ru and Os）when subjected to pressure exceeding 10 GPa \cite{IshimatsuJPSJ}. 
The ferromagnetic hcp Co changes to a nonmagnetic fcc phase（matching Rh and Ir）at extraordinary high pressure exceeding 150 GPa \cite{Yoo, IshimatsuPRB}. 
However, Ni remains in the fcc phase（matching Pd and Pt）irrespective of the high pressure applied, as expected \cite{IshimatsuJPSJ}.

Some metastable structures such as fcc Fe \cite{Pescia}, bcc Co \cite{Prinz}, fcc Co \cite{Schneider}, and bcc Ni \cite{Tian} have been realized using thin films epitaxially grown onto suitable substrates, 
all of which remain ferromagnetic at relatively low temperatures. 
These phases are stabilized within critical thicknesses, 
above which thermodynamically stable structures in bulk metals coexist. 
The realization of bcc Co and bcc Ni phases is very important because they do not appear under any (positive) pressures in bulk. 
In the case of thin films, interdiffusion and hybridization at the interface make data analysis and interpretation more complicated.

Phases differing from the bulk can also appear in a nanocrystalline form without support by a substrate. 
The resulting phase in nanocrystals (NCs) is determined by their preparation method and conditions, 
and is related to their size, shape, and surface morphology. 
Regarding 3$d$ magnetic metals, Co NCs have been found to crystallize into hcp, fcc, and $\beta$-Mn structures \cite{Sun, Puntes, Yang}, while Ni NCs exhibit fcc and hcp structures \cite{Jeon, Tzitzios, Han, Luo}. 
Interestingly, the hcp Ni NCs demonstrate a nonmagnetic behavior, 
in contrast to the retention of ferromagnetism by the three Co phases. 
On the other hand, no different structures have been observed in Fe NCs except ones grown under unusual conditions \cite{Peng, Wei, Bianco}. 
Although intensive studies have been devoted to metal NCs, the mechanisms that produce these different structures for specific nanocrystalline metals have not been fully elucidated. 

In this study, we synthesized hcp and fcc Ni NCs and investigated their structures, electronic states, and magnetic properties. 
X-ray diffraction (XRD) measurements, combined with heat treatment and application of high pressure, 
revealed the thermodynamic stability of the hcp phase exhibiting a 20\% volume expansion.  
Polarization dependent hard x-ray photoemission spectroscopy (HAXPES) indicated an electron transfer between valence bands 
resulting in a total band-structure energy reduction for the hcp phase in comparison with the fcc phase.  
Magnetization measurements using a superconducting quantum interference device (SQUID) magnetometer showed that the fcc phase is ferromagnetic while the hcp phase is nonmagnetic. 
Contemplating these results, we discuss the mechanisms which determine the thermodynamically stable crystal structure of Ni NCs.

\begin{figure}
\begin{center}
\includegraphics[scale=0.8]{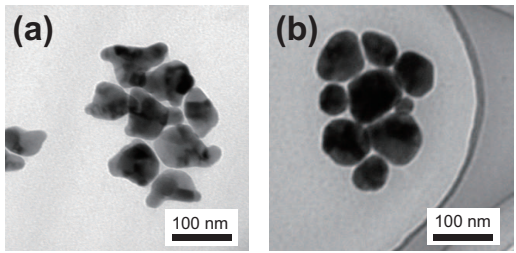}
\end{center}
\caption{Transmission electron microscopy images of (a) hexagonal close-packed (hcp) and (b) face-centered cubic (fcc) Ni nanocrystals (NCs).}
\label{f1}
\end{figure}

\begin{figure}
\begin{center}
\includegraphics[scale=0.35]{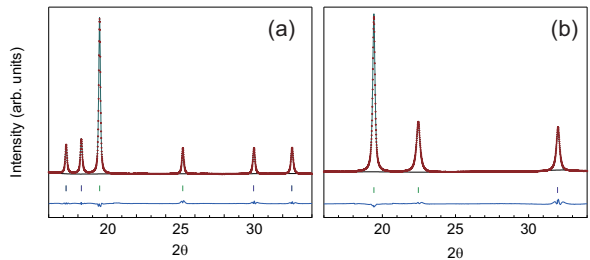}
\end{center}
\caption{Experimentally observed (data points marked with crosses, measured at RT) and calculated (solid curves) x-ray diffraction (XRD) patterns for the (a) hcp and (b) fcc Ni NCs.
}
\label{f2}
\end{figure}

\begin{table}
\caption{Refined structural parameters and reliability factors for the hcp and fcc Ni NCs at RT. 
The lattice constant for fcc Ni bulk is also shown \cite{Neighbours}. 
}
\begin{ruledtabular}			
\begin{tabular}{cccc}
	
			\     & hcp NCs & fcc NCs & fcc bulk \cite{Neighbours} \\ 
		 \hline
          $a_{hcp}$ (\AA) & 2.66080(13) & & \\
          $c_{hcp}$ (\AA) & 4.3465(4) &  &  \\
          $c/a$ (\AA) & 1.6335(2) & & \\			 
          $V_{hcp}$ (\AA$^3$) & 26.650(4) & & \\
          $a_{fcc}$ (\AA) & & 3.53743(8) & 3.5240(5) \\
          $V_{fcc}$ (\AA$^3$) & & 44.2654(18) & 43.76(3) \\
          $V\rm{/atom}$ (\AA$^3$) & 13.325(2) & 11.0664(5) & 10.941(5) \\
          $R_{\rm{wp}}$ (\%) & 2.785 & 2.852 &  \\
          $R_{\rm{I}}$ (\%) & 3.051 & 2.841 & \\
\end{tabular}
\label{Table.1}
\end{ruledtabular}
\end{table}

We obtained two different Ni NCs exhibiting hcp and fcc structures using organic phase synthesis. 
In the preparation of the hcp NCs, 
Nickel (II) acetylacetonate (1 mmol) was mixed in a reaction flask with oleylamine (3.5 ml) and 1-octadecene (5.5 ml).
Then, the air was pumped out, 
and the flask was heated at 120 $^\circ$C for 30 min with continuous pumping. 
Thereafter, the mixture was exposed to H$_2$(5\%)/Ar(95\%) gas, 
and the reaction temperature was raised to 240 $^\circ$C for a further 30 min. 
The resulting solution was cooled down to room temperature (RT),  
then the product was washed several times using a mixture of hexane and ethanol. 
The fcc NCs were obtained via the same procedure except only oleylamine (10 ml) was used as an organic solvent. 
Figures 1(a) and 1(b) show transmission electron microscopy images of the hcp and fcc NCs,  
both with crystal sizes in the 40-120 nm range. 
The shape of the hcp NCs suggests that the crystal growth is not isotropic, 
but instead has three equivalently promoted directions.

Figures 2(a) and 2(b) show XRD patterns for the hcp and fcc Ni NCs at RT; 
these data were collected with the BL12B2 beamline at SPring-8 with an incident x-ray wavelength of 0.6889 \AA. 
The Rietveld refinement of the XRD patterns was executed using the Rietan-FP program \cite{Izumi}. 
Each pattern was fitted with hcp and fcc structures without any impurities. 
The refined structure parameters and their reliability factors are listed in Table I. 
Although both structures are close-packed, 
their atomic packing densities are different from each other. 
The volume per atom of the hcp Ni NCs is more than 20\% larger than that of the fcc Ni bulk \cite{Neighbours} 
and the hcp Co bulk \cite{Taylor}, 
while the lattice constant of the fcc Ni NCs is almost equal to that of the fcc Ni bulk. 
This volume expansion of the hcp Ni phase has been observed in previous studies \cite{Jeon, Tzitzios, Luo}. 
The hcp Ni NCs with the large lattice constants have an axial ratio $c$/$a$ close to the ideal value of 1.633,  
indicating that the volume expansion is isotropic. 
Additionally, the peak widths in the XRD patterns of the hcp Ni NCs appear to be narrower than those of the fcc Ni NCs. 
Consequently, one may expect that the hcp phase is more stable than the fcc phase of the Ni NCs. 
To confirm this expectation, 
we performed post-annealing of the fcc Ni NCs at 330 $^\circ$C for 3 h in a mixture of oleylamine (20 ml) and oleic acid (20 ml). 
The stirrer was set to 900 rpm to prevent  aggregation of the NCs. 
The XRD pattern of the annealed sample showed the predominantly existing hcp phase, 
as well as some minor residual fcc phase. 
These results suggest that the hcp phase acquires thermodynamic stability by the large volume expansion.

\begin{figure*}
\begin{center}
\includegraphics[scale=0.5]{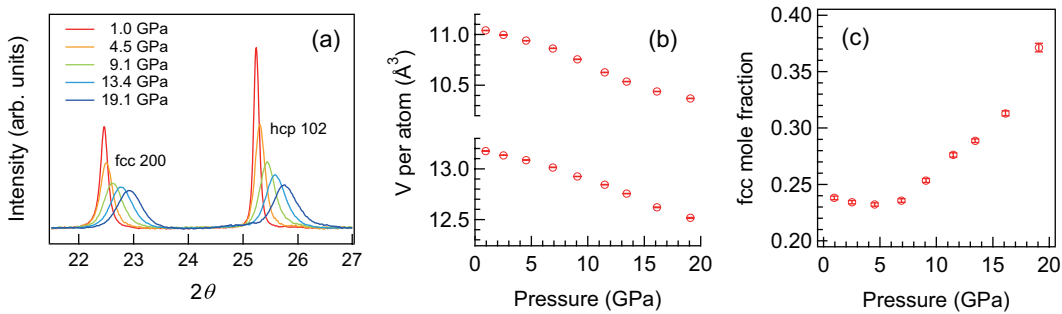}
\end{center}
\caption{Pressure dependence of (a) XRD patterns, (b) the volume per atom of the hcp (upper plot) and fcc (lower plot) phases, and (c) the mole fraction of the fcc phase for the annealed Ni NCs measured at RT.
}
\label{f3}
\end{figure*}

To clarify whether the stabilization of the hcp phase derives from the volume expansion corresponding to ``negative pressure'',
we investigated the pressure dependence of the composite ratio between the hcp and fcc phases for the annealed Ni NCs. 
A Mao-Bell-type diamond anvil cell was used to apply pressure 
which was transmitted through the medium of silicone oil,  
and measured by the line shift of ruby luminescence. 
Figure 3 (a) shows synchrotron XRD patterns of the annealed NCs under the application of positive pressure. 
The intensity of the fcc (200) is lower than that of the hcp (102) at lower pressures, 
but they become approximately equal at higher pressures. 
The Rietveld analysis showed that the volume per atom decreases with increasing pressure for both phases, as shown in Fig. 3(b),  
whereas the mole fraction of the fcc (hcp) phase increases (decreases) steadily from $\sim$5 GPa, as shown in Fig. 3(c). 
These results suggest that the volume compression increases the total energy of the hcp phase to be equivalent to that of the fcc phase. 
It is thus concluded that the large volume expansion plays an important role in the stabilization of the hcp phase.

\begin{figure}
\begin{center}
\includegraphics[scale=0.42]{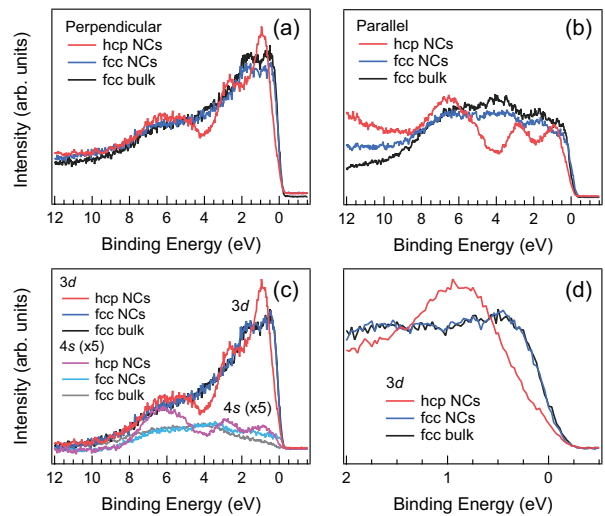}
\end{center}
\caption{Polarization-dependent valence-band hard x-ray photoemission spectroscopy (HAXPES) spectra of the hcp and fcc Ni NCs and fcc Ni bulk in the (a) perpendicular and (b) parallel geometries. 
(c) 3$d$ and 4$s$ spectra (including a minor portion of 4$p$) PDOS of the three samples. 
The 3$d$ spectrum was obtained by the subtraction of a Shirley background from the spectrum in the perpendicular geometry. 
The 4$s$ spectrum was obtained by the subtraction of the 3$d$ spectrum from the properly scaled spectrum in the parallel geometry. 
(d) 3$d$ PDOS near the Fermi level for the hcp and fcc Ni NCs and fcc Ni bulk.
}
\label{f4}
\end{figure}

The most important factor for determining the crystal structure of a transition metal is the $d$-occupation number $n_d$ \cite{Skriver, Pettifor}. 
Polarization dependent photoemission spectroscopy is a powerful technique to investigate the $d$ and $s$ partial density of states (PDOS) in a transition metal \cite{Sekiyama, Ueda, Weinen, Takegami}. 
Furthermore, the use of hard x-rays allows us to observe bulk electronic states \cite{Panaccione}. 
Valence band HAXPES measurements were thus performed with the BL12XU beamline at SPring-8. 
Two detectors were installed at positions perpendicular and parallel to the polarization of the incident beam \cite{Weinen, Takegami}. 
The photon energy was set to 6670 eV. 
For the following discussions, we employ semi-empirically estimated relative cross-sections of Ni 3$d$, 4$s$, and 4$p$ orbitals for polarized x-rays with 6 keV of energy: 
1.000, 27.87, 1.261 for the $parallel$ geometry, and 0.5656, 0.5835, 0.3010 for the $perpendicular$ geometry, respectively \cite{Ueda}. 
Additionally, the 3$d$ PDOS is much (10 to 20 times) larger than the 4$s$ and 4$p$ PDOS   
which are at the same level \cite{Papaconstantopoulos}. 
Therefore, the $perpendicular$ spectrum can be regarded as the 3$d$ PDOS, 
while the $parallel$ one should consist of the weighted summation of the 3$d$ and 4$s$ PDOS, 
and contributions of the 4$p$ PDOS to the spectra are negligible.

Figures 4(a) and 4(b) show valence band HAXPES spectra of the hcp and fcc Ni NCs and a fcc bulk (Ni plate, 99.9\%, Nilaco Co.), 
measured in the $perpendicular$ and $parallel$ geometries at RT. 
In each geometry, the two fcc spectra are alike, 
while significantly different from the hcp spectrum. 
The background was then subtracted from each spectrum by the Shirley method \cite{COMPRO}. 
To extract the 4$s$ PDOS, a properly scaled $perpendicular$ spectrum was subtracted from the $parallel$ spectrum  
so that the features seen in the $perpendicular$ spectrum were eliminated in the resulting one without forming new extrinsic structures. 
If the scaled $perpendicular$ spectrum is assumed to be the 3$d$ PDOS, 
the 4$s$ PDOS can be obtained by dividing the resulting $parallel$ spectrum by 27.87---the 3$d$ and 4$s$ PDOSs deductively obtained for the three samples are shown in Fig. 4(c). 
The PDOS of the fcc NCs and the fcc bulk are basically identical to the corresponding ones of previously reported HAXPES results on a fcc Ni metal \cite{Ueda}.

The PDOS of the hcp NCs qualitatively resemble the ones obtained by a self-consistent augmented plane-wave (APW) calculation for hcp Ni \cite{Papaconstantopoulos}. 
The calculated 3$d$ PDOS shows four characteristic structures: 
a large, sharp peak at the $E_F$;  
two sharp peaks below the $E_F$; 
and a broad peak at the lowest energy. 
The calculated 4$s$ and 4$p$ PDOS have no peaks at the $E_F$, 
but have three other structures almost at the same energies owing to hybridization. 
It is distinctive to the 4$s$ PDOS that the two peaks below the $E_F$ are broad and much smaller than the lowest energy peak. 
Our experimental results qualitatively reproduce these features if the $E_F$ of our hcp NCs is assumed to be about 1eV lower than that in the calculation.

The APW calculation assumed equal volume per atom as in the thermodynamically stable fcc bulk Ni (lattice constants $a$ = 2.497 \AA, $c$ = 4.044 \AA). 
However, our hcp Ni NCs have 20\% expanded volume per atom corresponding to a negative pressure, 
which causes upward energy shift of the 3$d$ band relative to the others 
because $d$ electrons would prefer smaller volume per atom than $s$ and $p$ electrons \cite{Miedema}. 
This transfers the valence electrons from the 3$d$ to the 4$s$ and 4$p$ bands 
and eliminates the large, sharp peak at the $E_F$ of the 3$d$ PDOS. 
The separations between the three characteristic peaks seem to be enlarged by this band shift and hybridization.

The 4$s$ PDOS shown in Fig. 4(c) has a line shape qualitatively consistent with the calculation,  
but the two peaks below the $E_F$ are somewhat larger. 
This is likely an indication of the finite inclusion of the 4$p$ PDOS. 
The calculation results showed that the two peaks of the 4$p$ PDOS are much (6 to 7 times) larger than those of the 4$s$ PDOS \cite{Papaconstantopoulos}. 
This is probably why there is a finite contribution of the 4$p$ electrons, 
although the relative cross-sections of 4$s$ and 4$p$ for the $parallel$ geometry are 27.87 and 1.261 \cite{Ueda}.

We now roughly estimate $n_d$ by tentatively ignoring the existence of 4$p$ electrons. 
Taking into consideration that the number of valence electrons per Ni atom is ten, 
we obtained $n_d$ and the number of 4$s$ electrons ($n_s$) for the three samples by integrating the PDOS shown in Fig. 4(c) 
and using the relation $n_d$ + $n_s$ =10. 
The $n_d$ value deduced for the fcc Ni bulk of 9.53 $\pm$ 0.04 is close to the value in the text book \cite{Kittel}. 
The same $n_d$ value is obtained for the fcc NCs, 
which is consistent with the equivalence of lattice constants (Table I). 
As is expected by the observation of the volume expansion of the hcp phase, 
the reduction in $n_d$ (9.39 $\pm$ 0.11) is confirmed to occur for the hcp NCs. 
Note that the real $n_d$ value, considering the 4$p$ electrons, should become smaller. 
The calculated DOS of Ni (both hcp \cite{Papaconstantopoulos} and fcc \cite{Vargas}), 
assuming the same volume per atom as the fcc Ni bulk, 
has a 4$p$ PDOS comparable to the 4$s$ PDOS below the $E_F$, 
and a greater 4$p$ PDOS than the 4$s$ PDOS above the $E_F$. 
Accordingly, with the volume expansion, the hcp Ni NCs should have a larger number of 4$p$ electrons ($n_p$) than the $n_s$. 
If we assume $n_p : n_s = 1.5 : 1$, the deduced $n_d$ value is 8.93 $\pm$ 0.08 for the fcc NCs and 8.61 $\pm$ 0.21 for the hcp NCs. 
Ignoring the 4$p$ electrons again, 
the centers of gravity of the valence bands, 
which correspond to average binding energies per valence electron, 
are now estimated to be 3.48 eV for the hcp NCs and 3.23 eV for the fcc NCs. 
This indicates that the hcp NCs are certainly more stable than the fcc NCs in terms of the total one-electron band-structure energy.

In the  APW calculation \cite{Papaconstantopoulos}, 
the hcp Ni exhibits a large, sharp DOS peak at the $E_F$ and a ferromagnetic ground state is predicted. 
However, this peak is eliminated because of the volume expansion in our hcp NCs. 
Actually, the 3$d$ PDOS of the hcp NCs is significantly smaller at the $E_F$ compared with those of the fcc NCs and the fcc bulk, as shown in Fig. 4(d). 
Therefore, the observed DOS of the hcp NCs are not supposed to satisfy the Stoner criterion for itinerant ferromagnetism. 
Figures 5(a) and 5(b) show the magnetization curves of the hcp and fcc NCs at 300 K. 
As expected, the hcp NCs show very weak magnetization compared with the fcc NCs, and are thus concluded to be nonmagnetic. 
The magnetic moment of the fcc NCs is estimated to be 0.629 $\mu_{\rm B}$ per atom, 
which is slightly larger than 0.60 $\mu_{\rm B}$ per atom of fcc bulk \cite{Kittel}. 

Finally, we discuss why the hcp phase is stabilized only in the NC form. 
As we found above, the one-electron band-structure energy of the hcp Ni NCs is 0.25 eV lower than that of the fcc Ni NCs, 
which is caused by the 20\% volume expansion. 
This expansion should also increase the electrostatic potential energy of the ion lattice. 
Since the hcp phase and the fcc phase coexist in the annealed Ni NCs, 
the total energies of the hcp and fcc phases must be close. 
Consequently, the loss of the electrostatic energy should be comparable to the gain of the band energy (2.5 eV per atom) in the hcp phase;  
this balance of energy is likely influenced by the surface-to-volume ratio. 
The surface atoms have lower coordination than the bulk, 
and thus have higher charge density per bond to maintain the charge neutrality of the topmost layer. 
For transition metals with $n_d > 5$, 
the additionally localized charge in the surface bonds is antibonding, 
and hence the surface atoms are less bound \cite{Citrin}. 
Accordingly, the increase of the electrostatic potential energy by the lattice expansion is suppressed at the surface. 
When the size of the Ni crystal is large enough, 
the total energy of the hcp phase is considered to be slightly higher than that of the fcc phase. 
As the size decreases, the total energy is reduced because the suppression is enhanced by the high surface-to-volume ratio, 
and the hcp phase is stabilized in NCs smaller than the size where the energies of both phases are equivalent. 
It is observed in Fig. 1(a) that the hcp Ni NCs tend to have horned shapes, 
which increase the surface-to-volume ratio compared with round shapes of the fcc NCs, 
and thus seems to be advantageous for the stabilization of the hcp phase. 
For quantitative discussion, further investigations considering size dependence and theoretical works are required.

\begin{figure}
\begin{center}
\includegraphics[scale=0.4]{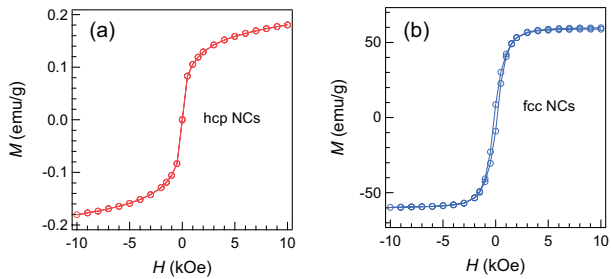}
\end{center}
\caption{Magnetization curves of (a) the hcp and (b) fcc Ni NCs at 300 K.}
\label{f5}
\end{figure}

In summary, we synthesized hcp and fcc Ni NCs with sizes in the 40-120 nm range. 
The hcp Ni phase is thermodynamically stable because of a 20\% volume expansion per atom 
that induces a $>$ 0.14 valence electron transfer decreasing $n_d$, 
a 0.25 eV band energy reduction per electron compared with the fcc Ni NCs, 
and an unusual nonmagnetic nature as well. 
The stabilization of the hcp Ni NCs can be understood from a viewpoint of the balance between the band energy gain and the electrostatic potential energy loss related to the large lattice expansion.

\begin{acknowledgments}
We would like to thank F. Okubo, N. Toyoda, Y. Nakagaki, T. Nishimura, Y. Yamada, S. Matsumoto, and Y. Hasegawa
for their help at an early stage of this study.
The XRD and HAXPES experiments at SPring-8  were performed with the approval of JASRI (Proposal Nos. 2017A4130, 2017B4255, 2017B4129, 2018A4142, 2019A4254, and 2019B4129).
This work was supported by a Japan Society for the Promotion of Science Grant-in-Aid for Scientific Research (C), No. 17K05501.
Transmission electron microscopy analyses were conducted with a JEM-1400 microscope spectrometer at the Analytical Research Center for Experimental Sciences, Saga University.
\end{acknowledgments}


\bibliography{basename of .bib file}

\end{document}